\documentclass[reprint,superscriptaddress,aps,prb,showpacs]{revtex4-1}
\usepackage{graphicx}
\usepackage{amsbsy,amssymb,amsmath,bm,ulem}

\graphicspath{{../fig/}}

\begin{document}

\title{Diversity of flux avalanche patterns in superconducting films}
\author{J. I. Vestg{\aa}rden}
\affiliation{Department of Physics, University of Oslo, P. O. Box
1048 Blindern, 0316 Oslo, Norway}
\author{D. V. Shantsev}
\affiliation{Department of Physics, University of Oslo, P. O. Box
1048 Blindern, 0316 Oslo, Norway}
\author{Y. M. Galperin}
\affiliation{Department of Physics, University of Oslo, P. O. Box
1048 Blindern, 0316 Oslo, Norway}
\affiliation{Physico-Technical Institute RAS, 194021 St. Petersburg, Russian Federation}
\author{T. H. Johansen}
\affiliation{Department of Physics, University of Oslo, P. O. Box
1048 Blindern, 0316 Oslo, Norway}
\affiliation{Institute for Superconducting and Electronic Materials,
University of Wollongong, Northfields Avenue, Wollongong, NSW 2522, Australia}

\begin{abstract}
The variety of morphologies in flux patterns created by 
thermomagnetic dendritic avalanches in type-II superconducting films is investigated 
using  numerical simulations. The avalanches are 
triggered by introducing a hot spot at the edge of a strip-shaped sample, which 
is initially prepared in a partially penetrated Bean critical state by slowly ramping
the transversely applied magnetic field.  The simulation scheme 
is based on a model accounting for the nonlinear and nonlocal 
electrodynamics of superconductors in the transverse geometry.
By systematically varying the parameters
representing the Joule heating, heat conduction in the film, and heat transfer 
to the substrate, a wide variety of avalanche patterns is formed, and 
 quantitative characterization of areal extension, branch width etc. is made.
The results show that branching is suppressed by the lateral heat diffusion, while
large Joule heating gives many branches, and
heat removal into the substrate limits the areal size. 
The morphology shows significant dependence also on
the initial flux penetration depth.

\end{abstract}

\pacs{74.25.Ha, 68.60.Dv,  74.78.-w }

\maketitle

\section{Introduction}
The gradual flux penetration in type-II superconducting films is
occasionally interrupted when large amounts of magnetic flux rush in
from the edges, forming complex dendritic patterns inside the
specimen. Experiments using magneto-optical imaging have revealed that 
such dendritic avalanches take place 
in numerous materials, e.g.,
Nb,\cite{duran95}
YBa$_2$Cu$_3$O$_{7-x}$,\cite{brull92,leiderer93,bolz00} 
MgB$_2$,\cite{johansen02,ye04}
Nb$_3$Sn,\cite{rudnev03} YNi$_2$B$_2$C,\cite{wimbush04}
Pb~\cite{vlasko-vlasov00} and NbN;\cite{rudnev05} 
see also Ref.~\onlinecite{altshuler04} for a review. The similarity in
the avalanche morphology in so different materials strongly suggests
that the origin of the phenomenon is of a universal nature. Indeed, it
is today widely agreed that the dendritic instability originates from
a thermomagnetic breakdown in the superconductor.\cite{mints81} This
can occur when a small temperature fluctuation locally weakens the
pinning of the vortices, causing some magnetic flux to advance into
the superconductor. Since moving flux releases heat, the local
temperature then increases further, and a positive feedback loop is
formed, which can lead to a rapid runaway in the temperature and
magnetic flux propagation.

In superconducting films placed in perpendicular magnetic fields, the
nonlocal electrodynamics complicates the theoretical description of
the instability. Thus, an analytical treatment was largely delayed
compared to the bulk case, and in particular, the conditions for
instability onset, and the origin of a fingering nature of the
avalanches in the film case were explained only relatively
recently.\cite{rakhmanov04,denisov05,aranson05} To follow the complete time
evolution of dendritic avalanches, including the cascades of branching
events, only numerical simulations have proved successful and indeed
produced very realistic
results.\cite{aranson05,vestgarden11,vestgarden12-sr} However, so far
little work has been done to systematically characterize these
patterns, and to our knowledge, no effort was made trying to identify
how various physical parameters influence the morphology of the
avalanches.

Such characterization is not readily done experimentally since in
practice it is difficult to vary the material parameters
independently.  Hence, the most feasible way to make a systematic
investigation is to carry out a numerical simulation study.  In this
paper we present results of such a systematic study of the morphology
of dendritic flux patterns resulting from the thermomagnetic
instability in superconducting films.  First, we identify the
dimensionless parameters that enter the governing equations.  Then,
avalanches are nucleated at an edge, where partially penetrated
critical-states serve as initial conditions, and we follow the
avalanche evolution until it ends in a frozen flux pattern.  The final
patterns created using different parameter values are compared,
analyzed quantitatively and discussed.

The paper is organized as follows: Section II describes the model and
introduces the key dimensionless parameters.  Section III reports 
and discusses the
results of our simulations where those parameters are systematically
varied. In addition, the dependence of the avalanche patters on the
initial critical-state is investigated. The results are summarized in Sec.~IV.

\section{Model}
\begin{figure}[tt]
  \centering
  \includegraphics[width=\columnwidth]{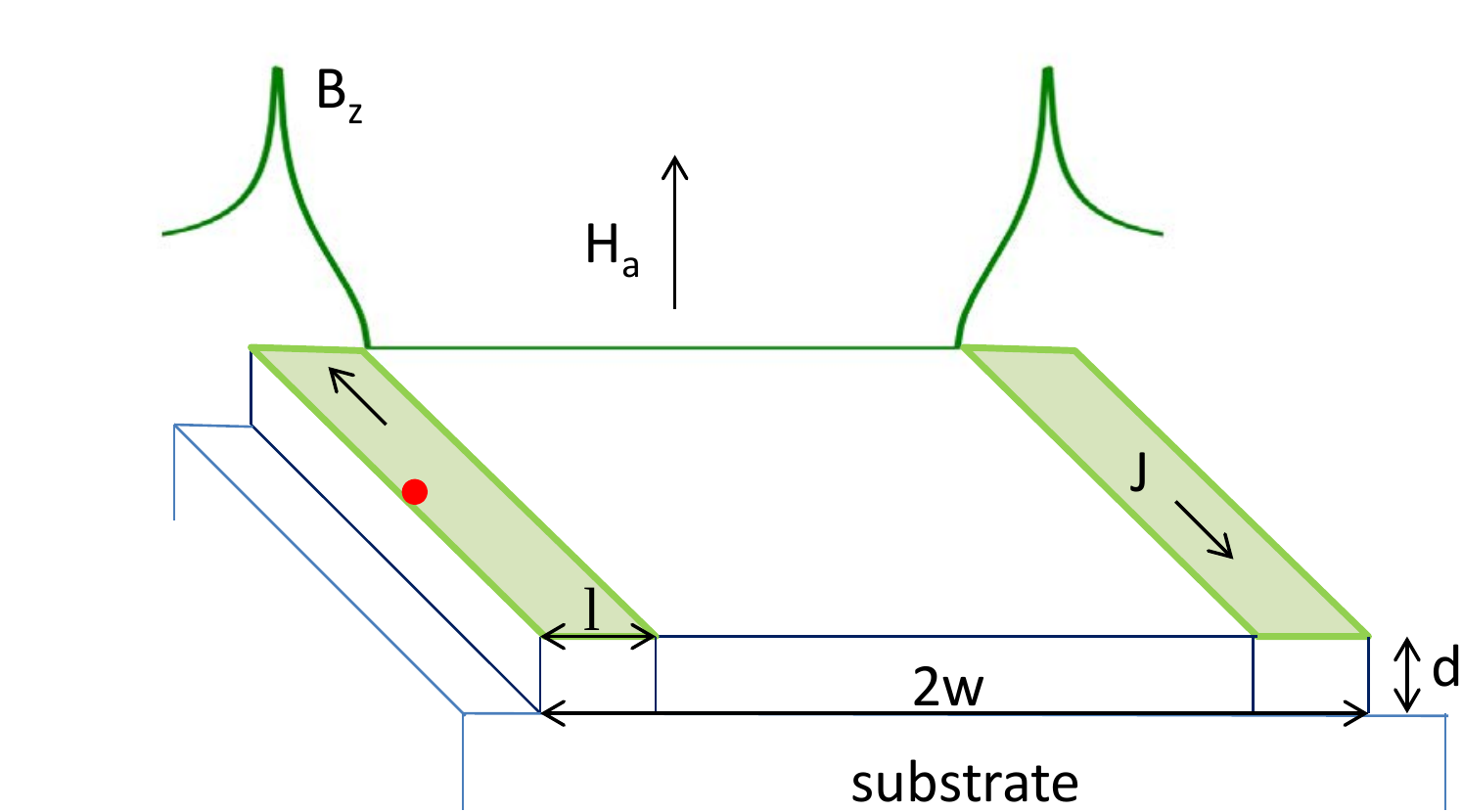}
  \caption{
    Sketch of the sample configuration.
    A long superconducting strip of thickness $d$ and half-width $w$, is 
    in thermal contact with a substrate. Over a region of width $l$ 
    near both edges a critical state with sheet current $J=J_c$
    and flux density $B_z$ exists as a hot-spot (red dot) is introduced to 
   trigger an avalanche.
    \label{fig1}}
\end{figure}

Consider a long superconducting strip of half-width $w$ and thickness
$d\ll w$, in thermal contact with a substrate, see Fig.~\ref{fig1}.
In the numerical simulations the following strategy is used: Starting
from a zero-field-cooled strip, we slowly ramp (increase) the applied
magnetic field $H_a$, while the thermal feedback is
turned off. This ensures that the flux penetration is gradual and the
temperature is everywhere equal to the substrate temperature $T_0$.
The spatial distributions of flux density $B_z$ and sheet current
$\mathbf J$ develop in agreement with the critical state model for the
transverse geometry.\cite{brandt93} The field ramp is stopped when the
flux penetration reaches the depth, $l$, where $l /w \sim 0.1$, and
that state is taken as initial condition for the upcoming avalanche.
The thermal feedback is then turned on, and a heat pulse is applied to
a small region near the edge, thus triggering the instability.  The
same protocol is repeated using different sample parameters, and for
each run we analyze the size and morphology of the flux distribution
frozen in the superconductor after the avalanche.

Before describing the physical model in detail, let us introduce the
units that will be used.  Time is expressed in units of
$t_0=\mu_0dw/\rho_n$, where $\mu_0$ is the vacuum magnetic
permeability and $\rho_n$ is the normal-state resistivity of the
superconductor. The sheet current $\mathbf J$ and magnetic field
$B_z/\mu_0$ are both expressed in units of the zero-temperature
critical sheet current, $J_{c0}=dj_{c0}$.  The temperature $T$ is
normalized to the critical value $T_c$, and the spatial coordinates
$(x,y)$ are in units of $w$.  Consequently, the Faraday law, $\dot
{\mathbf B}=-\nabla\times \mathbf E$, implies that the electric field
$\mathbf E$ is normalized by $\rho_nj_{c0}$.

We consider a superconductor described by a non-retarded relationship
between the electrical field and the current density as
follows:\cite{brandt95}
\begin{equation}
  \label{power-law-EJ}
  \mathbf E = \rho \mathbf J,~~
  \rho = 
  \begin{cases}
    1, & \text{if $J>J_c$ or $T>1$},\\
         (J/J_c)^{n-1}, 
& \text{otherwise}\, .
  \end{cases}
\end{equation}
Here $J \equiv |\mathbf{J}|$ while $n$ is the creep exponent. 
The temperature dependencies of $n$ and $J_c$ are taken as\cite{vestgarden11,ye04}
\begin{equation}
  \label{temperature-dependencies1}
\begin{split}  
  & J_c(T) = 1-T, \\
  & n(T) = n_1/T + n_0 \,   .
\end{split}
\end{equation}

The thermal properties of the superconducting film are specified by
the temperature dependence of the specific heat $c$, thermal
conductivity $\kappa$, and the coefficient of heat removal to the
substrate $h$.  We use here an approximation applied successfully in
previous modelling work, namely that all three parameters have cubic
$T$-dependencies.\cite{denisov06}

Based on this, the evolution of the local temperature is described by
the dimensionless diffusion equation,
\begin{equation}
   \label{dynamics-T}
   \dot T  = \alpha \nabla^2T -\beta(T-T_0)+\gamma T^{-3} \mathbf J\cdot \mathbf E
\   ,
\end{equation}
where $T_0$ is the normalized substrate temperature.
The coefficients
\begin{equation}
  \label{alpha-beta-gamma}
  \begin{split}
    \alpha=\frac{d}{w}\frac{\mu_0}{\rho_n}\frac{\kappa}{c},~~
    \beta =w\frac{\mu_0}{\rho_n}\frac{h}{c},~~
    \gamma=wd\frac{\mu_0}{c}\frac{j_{c0}^2}{T_c}\, ,
  \end{split}
\end{equation}
are also dimensionless, and involve $c$, $\kappa$ and $h$ evaluated at
$T_c$.  The parameter $\alpha$ has the meaning of a normalized
coefficient of thermal diffusion, and characterizes the smearing of
$T$ due to heat conduction within the film.  The $\beta $
characterizes the heat flow to the substrate, while $\gamma$ measures
the positive feedback due to Joule heating. For $\gamma=0$ the
evolution of the magnetic flux and temperature distributions are
decoupled and the film is always thermomagnetically stable.

Note that in this representation the parameter space is significantly
reduced compared to the dimensional description, since all the
$\kappa, h, c, T_c, j_{c0}$, and $\rho_n$ have been combined into
three dimensionless parameters.

To calculate the electromagnetic behavior we express the sheet current
through the local magnetization, $g$, as
\begin{equation}
  J_x = \partial g/\partial y,\quad   J_y = -\partial g/\partial x .
\end{equation}
Outside the sample, $g$ vanishes by definition. In an infinite or periodic space,
the Biot-Savart law for a thin film has a simple expression in the Fourier space. 
Inverting it and taking the time derivative yields\cite{vestgarden11}
\begin{equation}
  \dot g = \mathcal{F}^{-1}
  \left[ 
    \frac{2}{k}{\mathcal F}\left(\frac{\dot B_z}{\mu_0}-\dot H_a\right)
    \right],
  \label{dotg}
\end{equation}
where $\mathcal F$ and $\mathcal F^{-1}$ are the forward and inverse
Fourier transform, respectively, and $k=\sqrt{k_x^2+k_y^2}$ is the
in-plane wave-vector.  Inside the sample, the right-hand side of
Eq.~\eqref{dotg} is found from the Faraday law in combination with the
material law, Eq.~\eqref{power-law-EJ}.  Outside, $\dot B_z$ is found
implicitly by an iterative scheme requiring that $\dot g=0$. More
details about the simulation procedure are found in
Ref.~\onlinecite{vestgarden11}.

\begin{figure*}[p]
  \includegraphics[height=22cm]{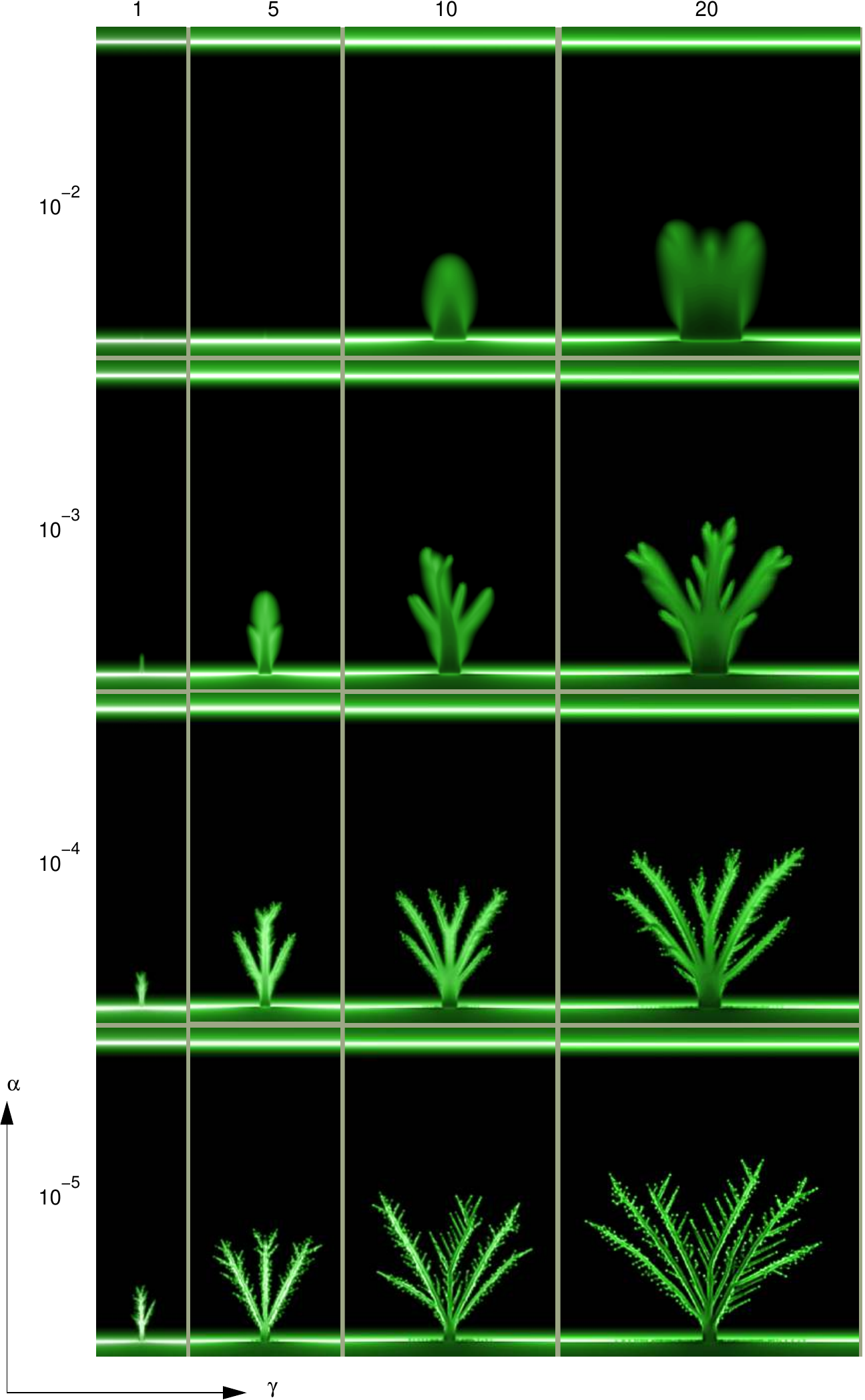} 
  \caption{
    \label{fig:morph1}
    Variety of avalanche flux distributions $B_z$ arranged in a matrix
    of panels with constant $\alpha$ in the rows, and with constant
    $\gamma$ in the columns. The image brightness represents the
    magnitude of $B_z$, as in experimental magneto-optical images, and
    the strip edges are seen as a pair of bright horizontal lines
    while the dark area between them is the flux-free Meissner state
    region of the superconducting strip.
  }
\end{figure*}

The sample occupies a $2\times 2$ square, with periodic boundary
conditions in one direction.  The other direction is extended with
extra space in order to take into account the magnetic field outside
the strip. The total area of the simulation covers a rectangle of size
$3\times 2$, which is discretized on a $768\times 512$ equidistant
grid.

The simulations were carried out using $T_0=0.15$, and the creep
exponent was set to $n=20/T-10$.  The initial state was prepared with
a ramp rate of $\dot H_a = 10^{-5}$, and the hot spot triggering the
avalanche had a size of $0.01\times 0.01$ (a cross consisting of 5
grid points) and a temperature of $T=1.5$.  During the avalanche the
field ramp was stopped.

\section{Results and Discussion}

The parameter space in these simulations is three-dimensional, and we
present and discuss first the various types of flux patterns produced
with $\beta=0.1 $, i.e., keeping a fixed coefficient of heat transfer
to the substrate. The initial flux penetration depth was $l = 0.11$
when a hot spot was introduced to trigger an avalanche.

\subsection{Varying $\alpha$ and $\gamma$}

Figure~2 displays a collection of final-state flux avalanche patterns obtained using 
values for  $\alpha$ and $\gamma$ in the range  $10^{-5} - 10^{-2}$ 
and  $1 - 20$,  respectively.
Evidently, within this part of parameter space one finds patterns which 
show striking similarities with 
those observed in experiments.\cite{duran95,brull92,leiderer93, bolz00,
johansen02,ye04,rudnev03,wimbush04,vlasko-vlasov00}
In addition, we find some type of patterns not previously reported.

The figure shows that with an increasing Joule heating, $\gamma$, and
a constant $\alpha$, the avalanches quickly become larger in size and
get more and more branches. At the same time, the branch width
remains essentially the same, although for increasing $\gamma$ the
main trunk near the edge gets steadily wider.

When instead keeping $\gamma$ constant and increasing the in-plane
heat diffusion, $\alpha$, one sees that the number of branches
decreases significantly. At the same time, the overall avalanche size
is not varying much. At maximum value, $\alpha=10^{-2}$, the
avalanches can hardly be characterized as dendritic, but rather as a
soft protrusion. Such non-dendritic shapes is a consequence of the thermal diffusion
being almost as fast as the electromagnetic propagation. 

The simulated $B_z$-maps reveal also several other interesting
features.  One is that one never finds branching to take place inside
the critical-state region near the edge.  Another is that even in the
largest avalanches, the branches rarely propagate past the sample
center.  Furthermore, branches never overlap, thus appearing to repel
each other.  Yet another is that the inner part of the branches most
often has the highest flux density, but in some cases they show a dark
low-$B_z$ core.  Note also that the flux density is always reduced in
the avalanche trunk as well as outside the sample, next to the root
location.
All these features correspond very well with experimental observations
using  magneto-optical  imaging, thus demonstrating a detailed 
correspondence between our simulations and reality.

\begin{figure}[b]
  \includegraphics[width=\columnwidth]{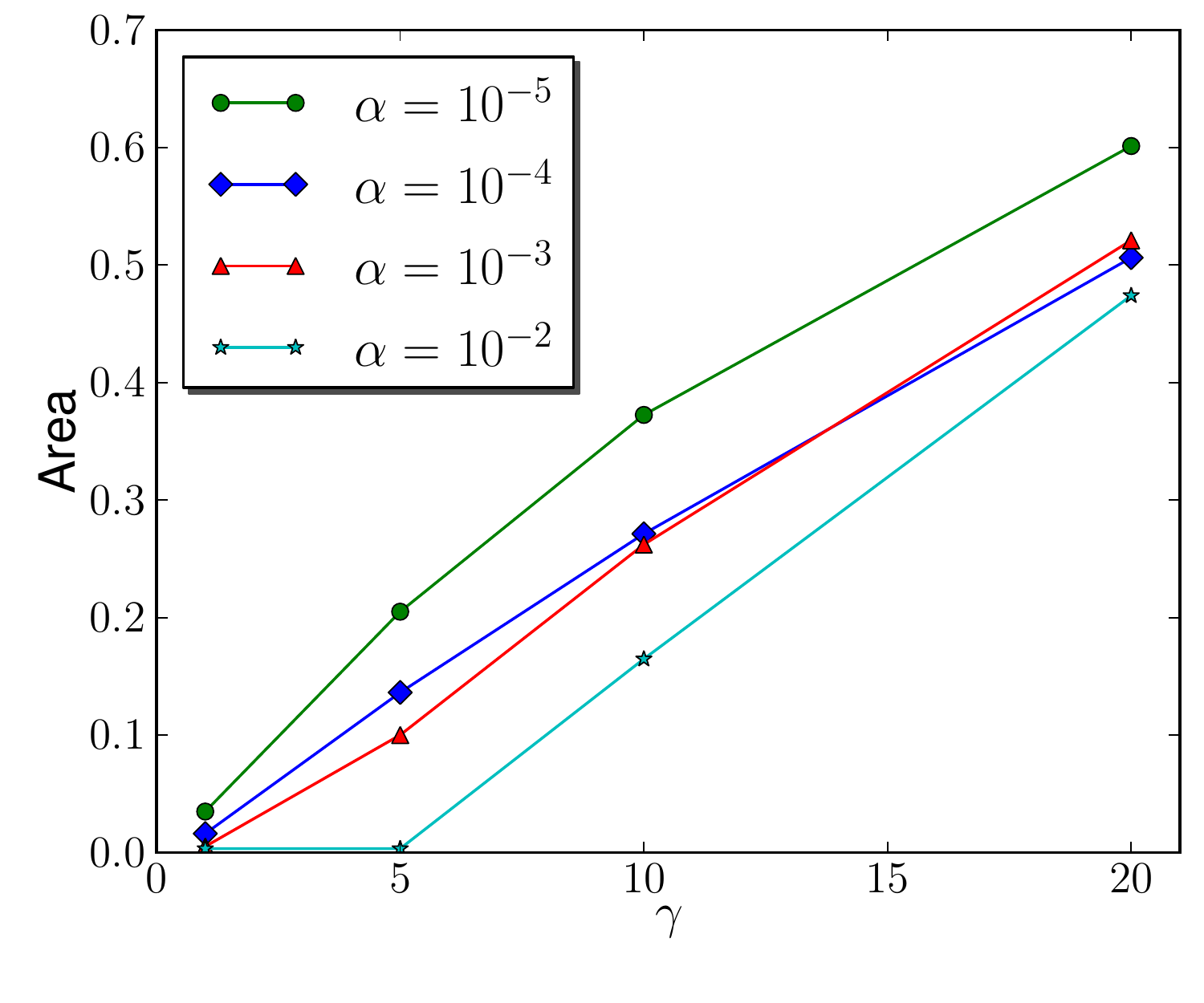} 
  \caption{
    \label{fig:area}
    Avalanche size in units of $ w^2$ as a function of $\gamma$ for different
    $\alpha$. The data are extracted from the flux patterns 
    displayed in the panels of Fig.~\ref{fig:morph1}.
  }
\end{figure}

A feature seen in Fig.~\ref{fig:morph1} which so far was not reported
neither from experiments nor from previous simulations, is the
appearance of quasi-periodic side branches in the panels with
small $\alpha$. The origin of these branches is not yet clear, but
a similar phenomenon is known from other systems, e.g.,
formation of dendrites in crystal growth.\cite{langer80} 

The panels  in the upper left corner of Fig.~\ref{fig:morph1},
corresponding to small $\gamma$ and large $\alpha$, do not show 
any visible trace of avalanche activity.
This indicates the existence of a region in parameter space
where the system is stable  towards even large perturbations, such 
as local heating above $T_c$. Hence, a guiding line for design of,
e.g., superconducting power-devices is to make materials and 
dimensions such that $\gamma$ is small and $\alpha$ is large. 
Note that the transition between stable and unstable behavior
is a crossover rather than a sharp phase boundary.

\begin{figure}[t]
  \includegraphics[width=\columnwidth]{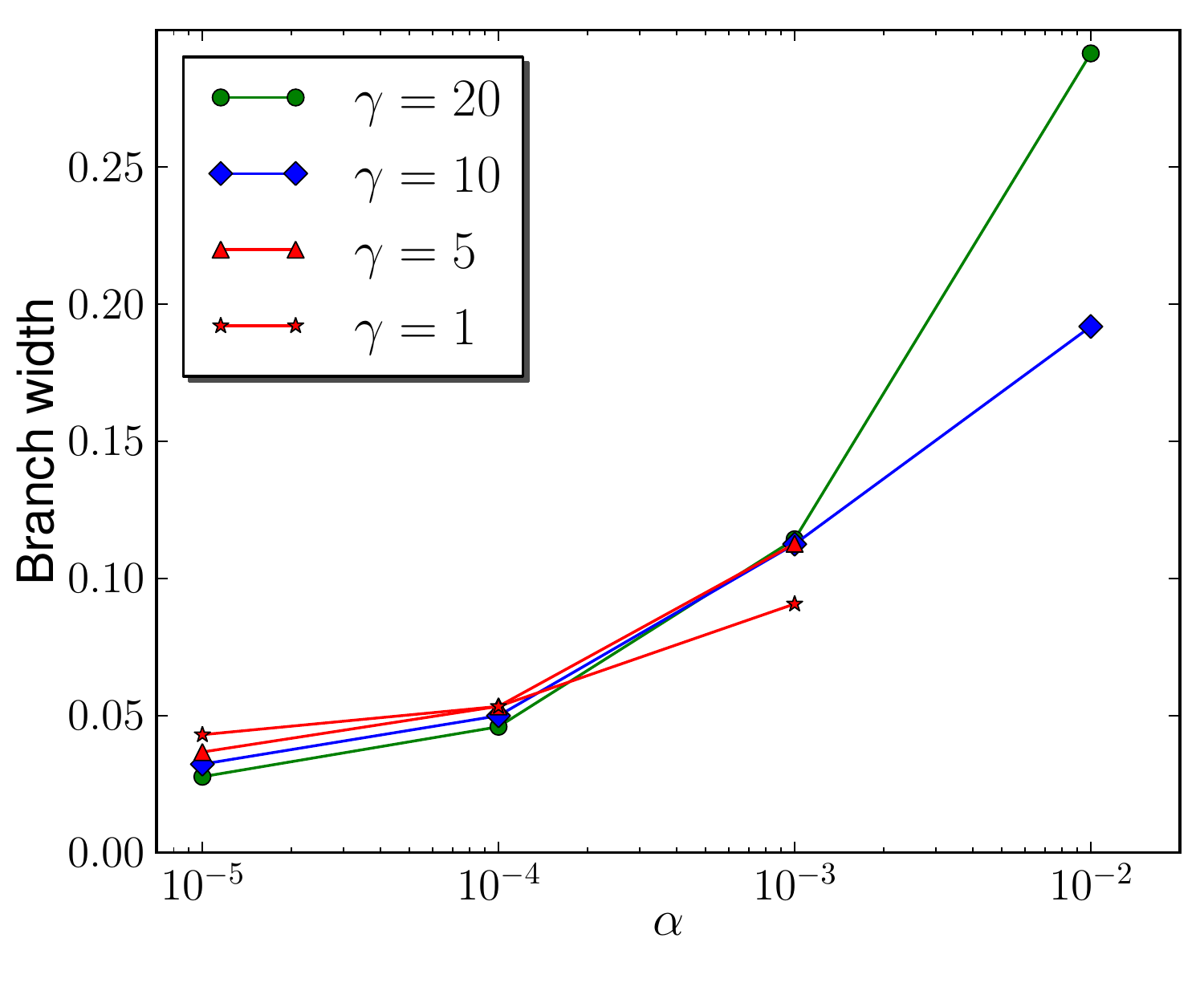}
  \caption{
    \label{fig:finger}
    Average branch width in units of $w$ as function of $\alpha$ 
    for various $\gamma$, extracted from the flux patterns
    shown in  Fig.~\ref{fig:morph1}.
  }
\end{figure}

Finally, note also that the flux structures in Fig.~\ref{fig:morph1} 
are more  symmetric and with straighter branches than the 
avalanche structures reported from previous simulation work
using the same formalism.\cite{vestgarden11,vestgarden12-sr}
This difference is due to the fact that in previous work
static disorder was introduced in the critical 
current density whereas here the sample was perfectly uniform. 
Regarding the role of disorder we therefore conclude that
 (i) Branching flux patterns may appear even in uniform materials,
in agreement with theoretical predictions
of the fingering instability in a uniform system.\cite{rakhmanov04,denisov05,aranson05}
(ii) It is
the non-linear and nonlocal electrodynamics, rather than spatial disorder, that is 
responsible for the branching of the flux structures.
(iii) The presence of disorder affects the exact path along which
the hot branches propagate. Straight main branches with quasi-periodic side branches
in a uniform sample become  wiggly and lose periodicity in a disordered sample.
(iv) Figure~\ref{fig:morph1} shows no sign of symmetric bifurcations, suggesting that
the process of symmetric tip splitting found in Ref.~\onlinecite{vestgarden12-sr} is also driven by disorder.

\begin{figure}[b]
  \includegraphics[width=\columnwidth]{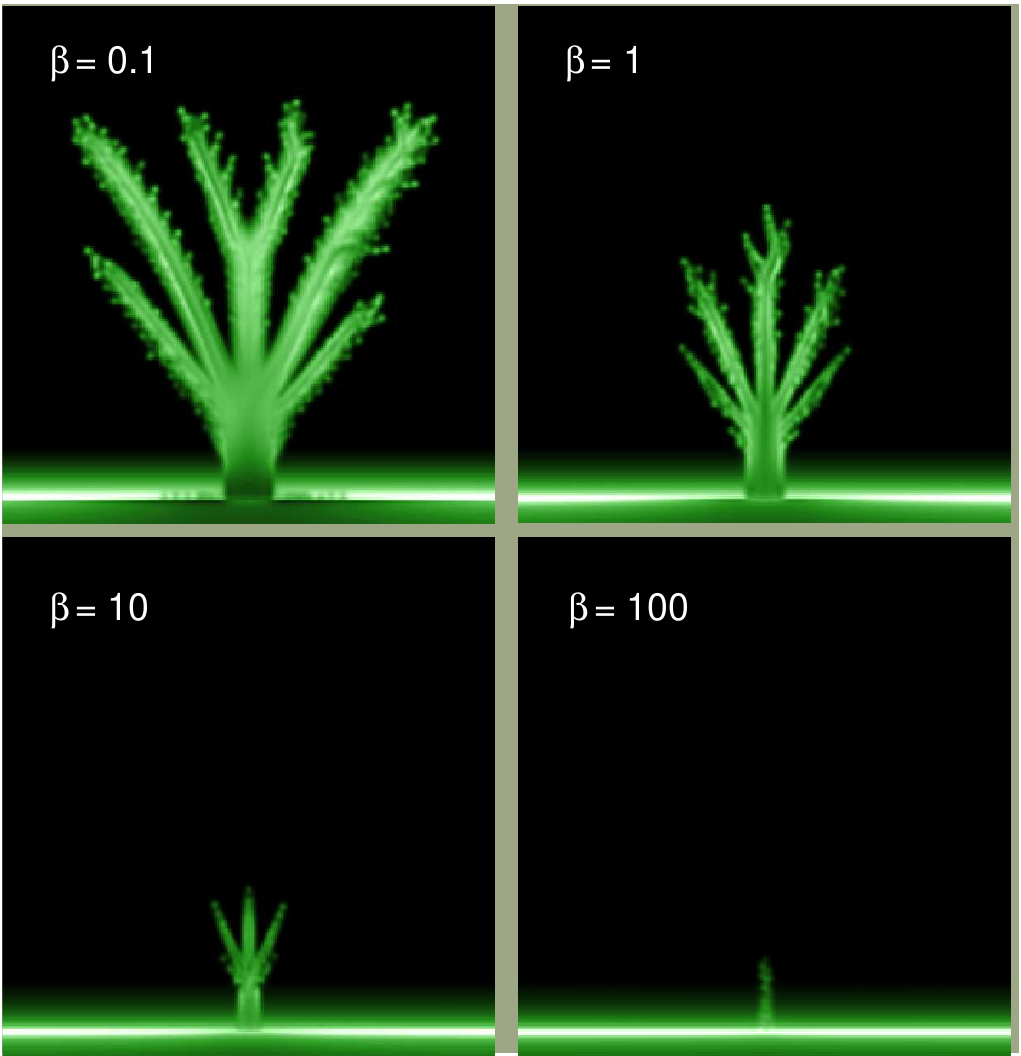} 
  \caption{
    \label{fig:vary-beta}
    Avalanche flux patterns  $B_z$  obtained  for various $\beta$,
    keeping constant  $\alpha = 10^{-4}$,  $\gamma = 10$
    and $l=0.11$. These panels show only
    the lower half of the strip.
  }
\end{figure}

\begin{figure}[b]
  \includegraphics[width=\columnwidth]{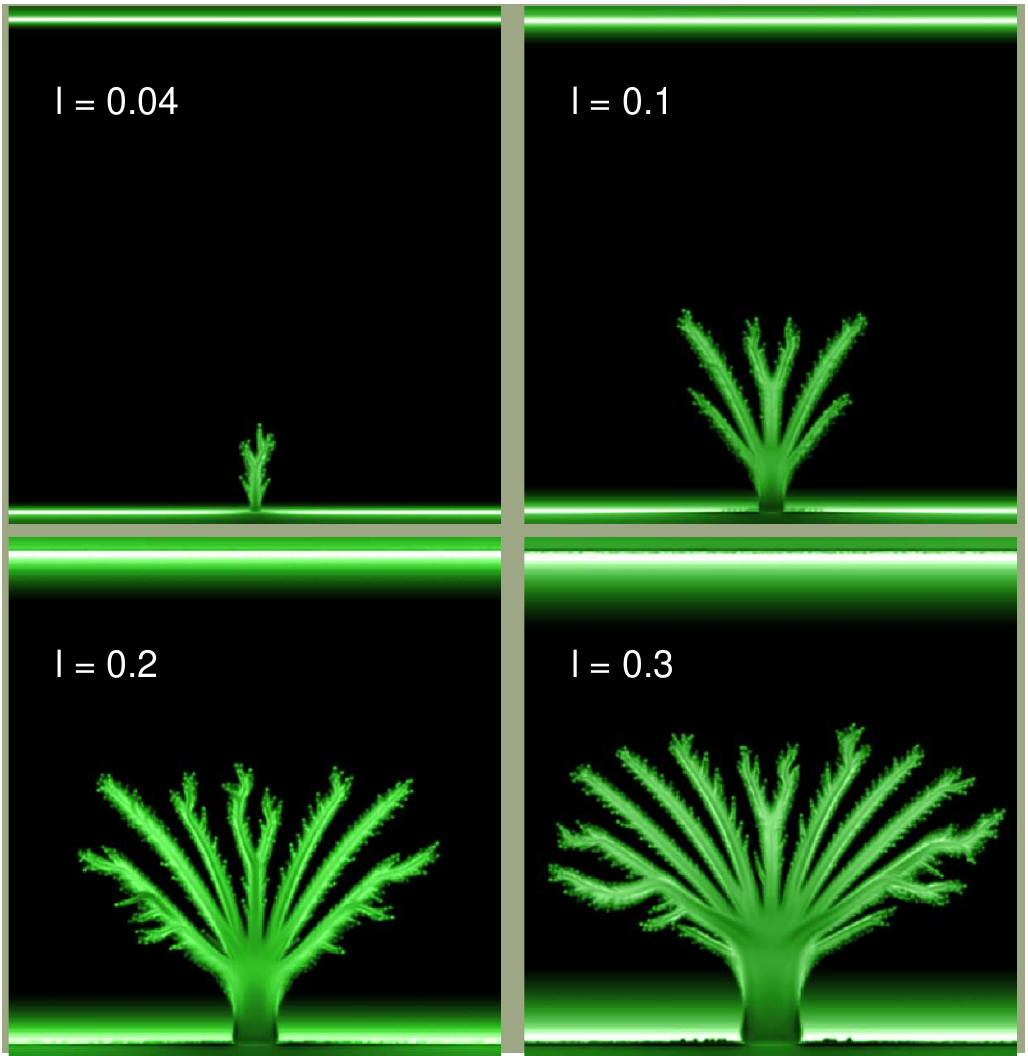} 
  \caption{
    Flux distributions $B_z$  for avalanches nucleated at different
    critical-state penetration depths $l=0.04$, 0.1, 0.2, and 0.3,
     with $\alpha = 10^{-4}$, $\beta = 0.1$ and $\gamma = 10$.
    \label{fig:vary-l}
  }
\end{figure}

As a quantitative measure of the effect of varying $\gamma$ and
$\alpha$, Fig.~\ref{fig:area} shows the total avalanche size, taken as
the area of the Meissner state region invaded by flux, plotted as a
function of $\gamma$ for various $\alpha$.  In spite that $\alpha$
spans 4 decades the graphs are not very different and increase almost
linearly with the same slope.  Evidently, the Joule heating parameter
$\gamma$ is a crucially important for the impact of an avalanche.

Presented in Fig.~\ref{fig:finger} is the average branch width plotted
as a function of $\alpha$ for various $\gamma$.  The width is
evaluated by dividing the area of the avalanche by the length of the
contour of $B_z=0$ around the structure.  The figure shows that the
width grows rapidly with increasing $\alpha$, and is quite weakly
depending on $\gamma$. This suggests that the width of the individual
branches in the dendritic structure to a large extent is controlled by
thermal diffusion.

\subsection{Varying $\beta$}

Consider next how the avalanche morphology is influenced by varying
the heat transfer to the substrate characterized by the parameter
$\beta$.  Figure~\ref{fig:vary-beta} illustrates the variety of flux
patterns obtained for $\beta$ ranging from 0.1 to 100, while keeping
constant the parameters $\alpha = 10^{-4}$, $\gamma = 10$, and
$l=0.11$.  It is evident that an increase in $\beta$ leads to
avalanches of smaller overall size and thinner branches, and to some
extent, also less degree of branching.  The behavior can be understood
from the role of $\beta$ in the heat diffusion equation,
Eq.~\eqref{dynamics-T}.  During an avalanche the term proportional to
$\beta$ represents a heat-sink that stabilizes the runaway and thus
tend to terminate the avalanche.\cite{vestgarden12-sr} A large heat
sink effect is also expected to give thinner branches.

\subsection{Varying $l$}

Finally, we consider how the avalanches depend on the penetration 
depth, $l$, of the critical-state background flux distribution.
Shown in Fig.~\ref{fig:vary-l} are four images of the strip 
where $l$ varies from 0.04 to 0.3 while
$\alpha$, $\beta$ and $\gamma$ are kept constant.
One sees that  large $l$ also gives large avalanches and many branches. 
This results is in agreement with previous experiment by 
Bolz et al.\cite{bolz00}

The images seen in  Fig.~\ref{fig:vary-l} show also a striking resemblance 
with previous experimental and simulation results at different 
substrate temperatures $T_0$.\cite{johansen02,vestgarden11} 
The results of this work suggests 
that the 
increasing avalanche size with $T_0$ is not because the avalanche propagation 
is sensitive to $T_0$, but rather because it is sensitive to the  
flux penetration depth prior to the avalanches,
which increases with $T_0$.

\section{Summary}

In this work we have used numerical simulations to investigate how
the dimensionless material parameters affect the size and morphologies
of dendritic flux avalanches in superconducting films.  The avalanches
were nucleated at the edge of a strip by a heat pulse and by varying
the parameters we have reproduced a wide range of morphologies
previously seen experimentally.  We find that increasing the
normalized coefficient of heat diffusion $\alpha$ gives fewer and
wider branches while strong heat removal to the substrate, quantified
by $\beta$, cause smaller total avalanche size and thinner branches.
Increasing values for the Joule heating parameter $\gamma$ gives more
branches and larger avalanches.  Parameter combinations with large
$\alpha$ and small $\gamma$ proved to be stable towards large
perturbations.
Quasi-periodic side branches were seen in the dendritic structures with small $\alpha$,
and their existence was attributed to the absence of spatial disorder. 
Finally, we found that the avalanche morphology is sensitive to the initial critical 
state before the avalanche, namely,
deeper initial penetration gives larger and more branching avalanches. 

For future work, it will be of interest to use the present dimensionless 
description to study how the linear stability diagram of superconductors 
is affected by varying the parameters $\alpha$, $\beta$, and $\gamma$.

\acknowledgments
The financial support from the Research Council of Norway is greatly 
acknowledged.


%

\end{document}